# The effect of the magnon dispersion on the longitudinal spin Seebeck effect in yttrium iron garnets (YIG)

Hyungyu Jin[1], Stephen R. Boona[1], Zihao Yang[2], Roberto C. Myers[2,3,4], and Joseph P. Heremans[1,3,4]

1. Department of Mechanical and Aerospace Engineering, The Ohio State University, Columbus, OH, 43210
2. Department of Electrical and Computer Engineering, The Ohio State University, Columbus, OH, 43210
3. Department of Materials Science and Engineering, The Ohio State University, Columbus, OH, 43210
4. Department of Physics, The Ohio State University, Columbus, OH, 43210

**ABSTRACT**

We study the temperature dependence of the longitudinal spin-Seebeck effect (LSSE) in a yttrium iron garnet $Y_3Fe_5O_{12}$ (YIG) / Pt system for samples of different thicknesses. In this system, the thermal spin torque is magnon-driven. The LSSE signal peaks at a specific temperature that depends on the YIG sample thickness. We also observe freeze-out of the LSSE signal at high magnetic fields, which we attribute to the opening of an energy gap in the magnon dispersion. We observe partial freeze-out of the LSSE signal even at room temperature, where $k_BT$ is much larger than the gap. This suggests that a subset of the magnon population with an energy below $k_BT_C$ ($T_C \sim 40$ K) contribute disproportionately to the LSSE; at temperatures below $T_C$, we label these magnons *subthermal magnons*. The *T*-dependence of the LSSE at temperatures below the maximum is interpreted in terms of a new empirical model that ascribes most of the temperature dependence to that of the thermally driven magnon flux.

# INTRODUCTION

The spin Seebeck effect (SSE), first reported in 2008[1], is the experimental manifestation of thermal spin transfer torque. The SSE[2,3] can be understood as the process by which a spin current is generated by a temperature gradient $\nabla T$ in one material, where it is carried by either magnons or spin-polarized electrons, and this spin current then crosses the interface into an adjacent normal metal (NM). Due to strong spin-orbit interactions in the NM, this injected spin current generates an electric field $E_{ISHE}$ via the inverse spin-Hall effect (ISHE). Since this electric field is related to $\nabla T$, it is possible in the regime of linear relations to define a spin-Seebeck coefficient $S_{SSE} = E_{ISHE} / \nabla T$. The effect manifests itself in two geometries, the transverse spin-Seebeck effect (TSSE)[1,4,5,6,7,8] and the longitudinal spin-Seebeck effect (LSSE)[9,10,11], in which the direction of the heat flux and of the spin flux are collinear and normal to the direction of the magnetization, respectively. In the latter geometry, a spin Peltier effect has been observed[12] to be the Onsager reciprocal of the LSSE. The potential issue of the admixture of a Nernst effect into the LSSE signal[13] has also been addressed.[14,15,16] The LSSE is well understood[17] at this point; however, it is only applied to ferromagnetic insulators such as yttrium iron garnet $Y_3Fe_5O_{12}$ (YIG), since the geometry of the measurement means that LSSE signals are indistinguishable from classical Nernst signals if ferromagnetic conductors are used. The general theory for LSSE is given in Ref [17], and the effect in the YIG/Pt system is attributed to magnon transport.[18,19,20,21] To explain this schematically, we can say that because there is a magnon contribution $\kappa_M$ to the thermal conductivity,[22,23] a magnon heat flux $\mathbf{j}_{QM} = -\kappa_M \nabla T$ exists in the presence of a temperature gradient, and thus there must also exist a flux of magnons, i.e., a spin or magnetization flux $\mathbf{j}_M$. In the simplest possible approximation, by treating the magnons as an ideal gas and ignoring the mode and frequency dependence of effects, one could even write that $\mathbf{j}_M = \mathbf{j}_{QM} (g_L \mu_B)/(k_B T) = -\kappa_M \nabla T (g_L \mu_B)/(k_B T)$. Here, $g_L$ is the Landé factor and $\mu_B$ the Bohr magneton. This thermally driven magnetization flux then crosses the interface through a process governed by a characteristic parameter known as the spin-mixing conductance $g_{\uparrow\downarrow}$, which is the ratio that relates the spin current generated in a material to the energy of the driving spin injection process.[24] The directional spin polarization vector is set by the magnetization **M** along the applied field **H** (as given in Fig. 1(b)). This overall spin polarization gives rise in the Pt metal to an inverse spin-Hall field given by $\mathbf{E}_{ISHE} = D_{ISHE} (\mathbf{j}_{QM} \times \mathbf{M})$, which is measured as a voltage $V_{ISHE} = |\mathbf{E}_{ISHE}| L$, where where $D_{ISHE}$ is the , and $L$ the length of the Pt strip.

The dependence of LSSE on YIG thickness has been studied in thin films,[25] along with the behavior at low[18] and high[26] temperatures and at high fields.[16] Particularly interesting is that near the Curie temperature, a specific temperature dependence has been observed that does not reflect that of the magnetization. Here, we present temperature-dependent LSSE data at cryogenic temperatures on both a bulk single crystal and a 4 μm thick film, and we show that they are different. This difference is interpreted in terms of the magnon thermal mean free path. We also extend the low temperature measurements to high magnetic fields, where we see that the LSSE signal is decreased at high field, which we interpret in terms of the magnon gap opening due to Zeeman splitting of the magnon dispersion. This suppression is also observed at 300 K, where the thermal energy is much larger than the magnon energy gap we open in our experiments. The fact that there still is a significant suppression of the LSSE at these temperatures and fields is taken as proof that low energy magnon modes appear to constitute a disproportionately large percentage of the LSSE signal at all temperatures. This leads us to divide the magnon population into two categories: *thermal magnons*, which we define as those magnons at all energies up to $T \times k_B$, where $T$ is the temperature of the experiment, and *subthermal magnons,* which we define as those at energies below $k_B T_C$. We define the cutoff temperature to be $T_C \sim 40$ K on the basis of the amount of freeze-out measured at room temperature, and note that the experimental magnon dispersion[27] at energies of $k_B T_C$ starts deviating significantly from a simple quadratic function of the wavevector. Obviously, the distinction between subthermal and thermal magnons only holds at $T>T_C$, and that below $T_C$ both populations are the same. The concept of subthermal magnons is inspired by the same concept used for phonons in the context of electron-phonon drag in semiconductors.[28] In phonon drag, the only phonons that can interact with the electrons are those with propagation vector limited to $2k_F$, where $k_F$ is the wavevector of the electrons at the Fermi surface. In semiconductors where $k_F$ is much smaller than the size of the Brillouin zone, and at high temperatures where the phonons of energy $k_B T$ have a much larger k-vector, this cutoff means that only those "subthermal" phonons with smaller k-vectors can participate in drag. Through our LSSE data we present here, we offer evidence that subthermal magnons in YIG interact more strongly with the electrons in Pt, similarly to phonon drag, such that LSSE is driven not equally by all modes of the magnon population at thermal equilibrium at any given temperature, but instead is biased toward low-energy subthermal magnons.

Since this manuscript was written, a manuscript has appeared on arXiv[29] which, for all intents and purposes, agrees very well with the data and conclusions we put forth here. In addition, we propose here an empirical model for the temperature dependence of the LSSE at low temperature and in thin films. The model takes into account the change of the magnon dispersion, from quadratic at the zone edge to kinear ("pseudo-acoutic" as defined by Plant[27]) at higher energies; this model explains the fact that the observed temperature dependence in thin films below 100 K does not actually follow a power law of $T$, but varies slowly from a $T^1$ law at the lowest temperatures, where the dispersion of the thermal magnons is mostly quadratic, to a $T^2$ law at higher temperatures.

# EXPERIMENTAL

Two different samples (A and B) were measured in this study. Sample A consists of a 4 μm thick single-crystalline YIG film grown on a 0.5 mm thick gadolinium gallium garnet (GGG) substrate with a 10 nm thick Pt layer deposited on top of the YIG. Sample B is a 0.5 mm thick single-crystalline YIG slab purchased from Princeton Scientific, Inc, upon which a 10 nm thick Pt layer was also deposited on one polished wide surface of the YIG slab using e-beam evaporation after careful cleaning. Dimensions ($L_x \times L_y \times L_z$, Fig. 1(b)) of the two samples are 2 mm × 6 mm × ~0.5 mm (sample A) and 5 mm × 10 mm × 0.5 mm (sample B).

For LSSE measurements, both samples were mounted in the same way as follows. First, two voltage leads were attached on two edges of the sample separated by $L_y$. Fine copper wires with 0.001 in. diameter were used for the voltage leads to minimize heat losses through the leads. Point contacts between the voltage leads and the sample were made using silver epoxy, and an effort was made to minimize the size of the contacts. The sample was sandwiched between two rectangular cubic boron nitride (c-BN) pads to promote uniform heat flux across the sample, and Apiezon N grease was used between the sample and c-BN pads to provide uniform thermal contacts. Three 120 Ω resistive heaters connected in series were attached on the top of the upper c-BN pad by silver epoxy. A Cernox thermometer was attached to each c-BN pad using Lakeshore VGE-7031 varnish. The whole sample block was then pushed onto the sample platform wherein the Apiezon N grease was used as a thermal contact. The sample block and the platform were wrapped together by an insulated wire in order to mechanically secure the heterostructure and provide a solid thermal contact between them. Fig. 1(a) shows a picture of sample A mounted in the way described above.

The LSSE measurements were performed using a liquid helium cryostat for sample A and a Physical Properties Measurement System (PPMS) for sample B. Sample B was also measured in the liquid helium cryostat for a cross-check, and the data obtained from the two different instruments showed good agreement in terms of the temperature dependence of LSSE signal. We observed that the magnitude of the signal can be slightly different for different measurement sets, possibly because of an error in the measured temperature difference induced by varying thermal contact resistances between the sample and the c-BN pads. Nevertheless, we confirmed that the temperature dependence itself is not affected by the small variability of the magnitude. The measurement configuration is shown in Fig. 1(b). First, the Cernox thermometers were calibrated as a function of temperature from room temperature down to 2 K. Resistances of the thermometers were measured at different temperatures using an AC bridge, and those values were interpolated to extract actual temperatures during LSSE measurements. LSSE measurements were made at different temperatures, and ~1h stabilization time was given after changing the base temperature to ensure thermal equilibrium between the sample and the system. After a temperature gradient $\nabla_z T$ was applied across the sample, additional 15 ~ 30 min wait time was used. The transverse voltage $V_y$ was measured using a Keithley 2182A nanovoltmeter while the magnetic field $H_x$ was continuously swept between a same $H_x$ value in both polarities and in both directions ($+H_x \rightarrow -H_x$ and $-H_x \rightarrow +H_x$). For low field measurement, $H_x$ was swept between ± 800 Oe at 4 Oe/s, and for high field measurement, between ± 70 kOe at 30 Oe/s. The temperature difference $\Delta T_z$ was separately measured from the attached thermometers using the same heater power as that used for the $V_y$ measurement. The sample chamber was kept under a high vacuum (<$10^{-6}$ torr) and covered by the gold-plated radiation shield to minimize convective and radiative heat losses, respectively.

# RESULTS AND DISCUSSION

Raw traces of the measured transverse voltage $V_y$ as a function of applied field $H_x$ for both samples are shown in Fig. 2. Around room temperature, both samples show clear LSSE signals under an applied temperature gradient $\nabla T_z$, wherein $V_y$ switches sign with the magnetization in the hysteresis loops (Fig 2(a), (d)). The magnitude of the switch, $\Delta V_y$, is defined as the difference between the saturation voltages.[7] The signals become noisier at low temperatures (Fig. 2(b), (e)) due to the smaller magnitude of both the voltage signals and the temperature differences across the samples. The raw traces of sample B exhibit an additional loop at low $H_x$ values, and the size of this loop is almost independent of temperature. While some of the previous studies reported a similar repeatable trend in their LSSE signals,[9,30] no explicit description has been given thus far to explain these observations. Based on the lack of temperature dependence of the loop size, we may tentatively attribute the additional loop to the presence of magnetic domains in the bulk sample. Fig. 2(c) and 2(f) show that $\Delta V_y$ varies linearly with $\Delta T_z$ for both samples, so that a spin Seebeck coefficient $S_y \equiv E_y / \nabla T_z = L_z \Delta V_y / 2 L_y \Delta T_z$ can be defined.

The same LSSE measurement was extended up to $H_x = \pm 70$ kOe for sample B. At $T_{avg} = 296$ K, a gradual reduction of $V_y / \Delta T_z$ is observed as $H_x$ increases (Fig. 3(a)). This behavior around room temperature is similar to what has been reported by Kikkawa et al.[16] also on Pt/YIG. The authors attributed the decrease of the LSSE signals to the suppression of magnon excitation caused by the magnon gap opening under an applied $H$. At low energies ($T = \hbar \omega / k_B < 30$ K), the magnon dispersion relation for YIG can be expressed reasonably well using the classic Heisenberg ferromagnet model for spin waves,[31] which is given as[32]

$$\hbar \omega_k = \mu_B g_L B + D a^2 k^2 \quad (1),$$

where $\hbar$ is Planck's constant, $\omega_k$ is the angular frequency of the magnon mode with wave vector $k$, $\mu_B$ is the Bohr magneton, $g_L$ is the Landé factor ($g_L$ = 2.046 for YIG[33]), $D$ is a spin-wave stiffness parameter, and $a$ is the size of the unit cell. Applying a non-zero external $H$ creates a forbidden zone in the dispersion (Eq. (1)) as well as in the density of states of magnons due to the Zeeman effect,[31] thus freezing out magnons with energy $\hbar\omega < \mu_B g_L B$. Since thermally excited magnons are believed to induce the SSE in ferromagnetic insulators[18,19,20,21,25], their suppression by an external $H$ is expected to reduce the measured ISHE voltage. Fig. 3(b) shows that the suppression of $V_y / \Delta T_z$ in applied $H$ is more pronounced at $T_{avg}$ = 10 K. If we define $V_{y,max}$ as the maximum value of the ISHE voltage obtained at low $H$, the $V_y / \Delta T_z$ at $H_x = \pm$ 70 kOe ($V_{y,70kOe} / \Delta T_z$) is less than 50 % of $V_{y,max} / \Delta T_z$. Surprisingly, about 20% suppression is observed at all temperatures up to $T_{avg}$ = 296 K. We compared the magnitude of $V_{y,max} / \Delta T_z$ in these high $H$ measurements (Fig. 3(c) and (d)) with that in the independently measured low $H$ results on sample B (Fig. 2(d)-(f)), and found that they agree well with each other.

The significant suppression of the LSSE signal at high $H$ at $T_{avg}$ = 10 K hints at which magnons are mainly responsible for the LSSE in Pt/YIG, which is further revealed by investigating the temperature dependence of the suppression effect. We can define $\Delta S_y = S_{y,max} - S_{y,70kOe}$, where $S_{y,max} \equiv L_z \Delta V_{y,max} / 2 L_y \Delta T_z$ and $S_{y,70kOe} \equiv L_z \Delta V_{y,70kOe} / 2 L_y \Delta T_z$. Fig. 4(a) shows the temperature dependence of $\Delta S_y / S_{y,max}$ derived from the high $H$ measurements (Fig. 3) on sample B. Above $T_{avg}$ = 30 K, the $\Delta S_y / S_{y,max}$ is almost constant at ~ 0.2 (inset in Fig. 4(a)), indicating that about 20% of the LSSE signal is suppressed at $H_x = \pm$ 70 kOe. The $\Delta S_y / S_{y,max}$ increases gradually below 30 K and then shows a drastic increase below 15 K reaching ~ 0.9 at about 7 K, suggesting that the LSSE almost disappears at this temperature and field. For YIG, at $H_x = \pm$ 70 kOe, a gap of $\mu_B g_L H / k_B = 9.6$ K opens in the dispersion relation given by Eq. (1) as shown in Fig. 4(b). Therefore, it is likely that most magnons are suppressed below 8 K in $H_x = \pm 70$ kOe and so is the LSSE, which is consistent with the experimental result in Fig. 4(a). But, as had already been pointed out in Ref [23], magnon suppression cannot be complete in $H_x = \pm 70$ kOe fields at temperatures above 10 K. Some amount of field suppression of the LSSE is measured at all temperatures, while not in specific heat or thermal conductivity.[23] Here, we verified experimentally on another piece of Sample A that the thermal conductivity of YIG is independent of magnetic field at 300 K to within less than 1%, the experimental accuracy of that measurement. We thus conclude that low-energy magnons (i.e. those of energy below $\sim k_B \times T_C$) must play a more prominent role in producing the LSSE. We estimate the value of the critical temperature $T_C \sim 40$ K using Maxwell-Boltzmann statistics by setting $0.2 = \Delta S_y / S_{y,\max} = \exp(\mu_B g_L B / k_B T_C) - 1$, but note that there is considerable uncertainty in the precise energy scale of the relevant modes, since we expect all magnon modes to contribute to thermal spin transport and LSSE, just some to contribute more than others. In particular, the inflexion point observed in the low-temperature behavior of $\Delta S_y / S_{y,\max}$ at 15 K in Fig. 4(a) seems to single this temperature out as well. Similar to the case of phonons,[28,34] we call these low-energy ($E < k_B T_C$), long-wavelength magnons *subthermal magnons* to distinguish them from the *thermal magnons* with relatively higher energies and short wavelengths that populate the magnon spectrum in thermal equilibrium at $T > T_C$.

We now discuss the temperature and thickness dependence of the LSSE in Pt/YIG. Fig. 5 shows the temperature dependence of $S_y$ for sample A and sample B between 10 K and 300 K. The $S_y$ of sample A has a maximum at 180 K. The $S_y$ of sample B, meanwhile, has a maximum at 70 K and a $\sim T^{1.19}$ (dashed line) dependence at low temperature. The longitudinal thermal conductivity $\kappa_{xxx}$ was measured independently, but is not shown because the data reproduce those of Slack & Oliver[35] quite exactly. The temperature dependence of the resistivity of the Pt film on sample B was measured to change from 0.35 µΩ m below 5 K to 0.52 µΩ m at room temperature. The temperature dependence of both these properties is quite different from that of the observed LSSE signal, which suggests that there is likely no significant correlation between them. In the absence of information about the temperature dependence of the spin Hall angle in Pt and of the spin mixing conductance at the YIG/Pt interface, we concentrate on the temperature-dependence of the thermally driven magnon flux.

Therefore, we assume that the temperature dependence of the LSSE signal arises primarily from that of the thermally driven magnon flux that reaches the interface. This interpretation, supported by the results of Ref. [26], allows us to neglect the *T*-dependence of the ISHE, the spin-Hall angle in the Pt, and the spin-mixing conductance across the interface. In general, since the magnetization flux at the YIG/Pt interface is in YIG driven by a temperature gradient, we can describe this flux by

$$j_M = g_L \mu_B j_\# = g_L \mu_B L_{\#T} \nabla T \qquad (2),$$

where $j_\#$ is the magnon number flux, and $L_{\#T}$ is the effective kinetic coefficient that relates this number flux to the accelerating force, the temperature gradient. The number flux is also related to the magnon heat flux $j_Q$, which is itself related to the magnon thermal conductivity $\kappa_M$ by Fourier's law $j_Q = -\kappa_M \nabla T$. Using the Boltzman diffusion equation in the relaxation time ($\tau$) approximation, both kinetic coefficients can be written for each magnon mode as an integral over the magnon frequencies (or energy $E$):[36]

$$L_{\#T} = \int_0^{E_{MAX}} v_G^2 \tau(E) \mathcal{D}(E) \left(\frac{\partial f}{\partial T}\right) dE$$

$$\kappa_M = \frac{1}{3} \int_0^{E_{MAX}} E v_G^2 \tau(E) \mathcal{D}(E) \left(\frac{\partial f}{\partial T}\right) dE \qquad (3).$$

Here $\mathcal{D}(E)$ is the magnon density of states and $f$ is the Bose-Einstein statistical distribution function, so that $\frac{\partial f}{\partial T} = \frac{x}{T}\frac{e^x}{(e^x-1)^2}$ where $x \equiv E/k_B T$ is the reduced energy.

At very low temperature where the magnon dispersion is purely quadratic[27] and given by Eq. (1), the group velocity and density of states in the absence of magnetic field become:

$$v_G = \frac{2}{\hbar}\sqrt{Da^2}\, E^{1/2}$$

$$\mathcal{D}(E) = \frac{1}{4\pi^2}\frac{E^{1/2}}{(Da^2)^{3/2}} \qquad (4).$$

An experimental estimate of the magnon thermal mean free path $\ell_M$ is given by Ref. [23] for $T > 2$ K. In the ballistic regime where the magnon mean free path is limited by sample size, $l_M$ is temperature-independent and $\tau(E) = \ell_M / v_G$. Consequently, we can derive the kinetic coefficient of the magnon number flux as well as the thermal conductivity:

$$L_{\#T} = \frac{3\ell_M}{2\pi^2}\frac{k_B^2}{\hbar}\frac{1}{Da^2}T\int_0^\infty \frac{x^2 e^x}{(e^x-1)^2}dx$$

$$\kappa_M = \frac{\ell_M}{3\pi^2}\frac{k_B^3}{\hbar}\frac{1}{Da^2}T^2\int_0^\infty \frac{x^3 e^x}{(e^x-1)^2}dx \qquad (5).$$

One recognizes in Eq. (5) the familiar magnon thermal conductivity $T^2$ scaling law. In that case, Eq. (5) predicts that the thermally driven magnon flux, and thus the LSSE signal, should scale as $T^1$ (recall $\mathbf{j}_M = -\kappa_M \nabla T (g_L \mu_B)/(k_B T)$). At $T < 2$ K, the $T^2$ law for $\kappa_M$ is observed,[22] but not[23] above 2 K, for two reasons. First, because the measurements are taken on bulk samples, the ballistic regime is not reached until $T < 2$K and there is a dependence of $l_M$ on $T$. The second reason is the non-parabolicity of the magnon dispersions. Eq. (1) already breaks down at energies below 10 K: indeed Ref [27] shows that the temperature dependence of the saturation magnetization curves departs from the Bloch $T^{3/2}$ law already below 10K.

At higher energies (> 8 meV) and temperatures, the magnon dispersion becomes linear[27], and the thermal conductivity in the ballistic regime is expected to follow a $T^3$ law, with $L_{\#T} \propto T^2$. In terms of group velocity, it increases with energy when the dispersion is quadratic, but saturates at a maximum value $v_{GS}$ when the dispersion becomes linear. These features can be captured by fitting the measured "pseudo-acoustic" magnon dispersion curves in YIG[27] (and ignoring the pseudo-optical modes) with a phenomenological model for the magnon dispersion written by analogy with the electron dispersion in narrow gap semiconductors such as PbTe[37] as:

$$\gamma(E) \equiv E(1+\frac{E}{E_0}) = Da^2 k^2 \qquad (6),$$

where $E_0$ parametrizes the saturation value $v_{GS}$ of the group velocity $v_G$, as shown below.[38] The group velocity and DOS are given by:

$$v_G = \frac{2\sqrt{Da^2}}{\hbar} \frac{\sqrt{\gamma}}{\gamma'}$$

$$\mathcal{D}(E) = \frac{1}{4\pi^2} \frac{\gamma'\sqrt{\gamma}}{(Da^2)^{3/2}} \qquad (7).$$

$$\gamma' \equiv \frac{d\gamma}{dE} = 1 + 2\frac{E}{E_0}$$

The value of $E_0$ is derived from $v_{GS}$ by taking the limit $v_{GS} = \lim_{E\to\infty}(v_G) = \frac{\sqrt{DE_0 a^2}}{\hbar}$. Ref [27] reports for $D = 46$ K $\times k_B$, $v_{GS} = 1.8 \times 10^4$ m s$^{-1}$ along the (0,0,l) axes, and $E_0 = 270$ K $\times k_B$.

The integrals in Eq. (3) can be fully evaluated and expressed in terms of $\ell_M$:

$$L_{\#T} = \frac{1}{3k_B T^2} \int_0^\infty \ell_M E v_G \mathcal{D}(E) \frac{e^x}{(e^x - 1)^2} dE$$

$$\kappa_M = \frac{1}{3k_B T^2} \int_0^\infty \ell_M E^2 v_G \mathcal{D}(E) \frac{e^x}{(e^x - 1)^2} dE$$

(8).

In the ballistic regime with a constant $\ell_M$:

$$L_{\#T} = \frac{3\ell_M}{2\pi^2} \frac{k_B^2}{\hbar} \frac{1}{Da^2} T \int_0^\infty \left(1 + x \frac{k_B T}{E_0}\right) \frac{x^2 e^x}{(e^x - 1)^2} dx$$

$$\kappa_M = \frac{\ell_M}{3\pi^2} \frac{k_B^3}{\hbar} \frac{1}{Da^2} T^2 \int_0^\infty \left(1 + x \frac{k_B T}{E_0}\right) \frac{x^3 e^x}{(e^x - 1)^2} dx$$

(9).

From Eqs (5) and (9), a temperature-dependence of the LSSE signal intermediate between $T^2$ and $T^1$ is expected, which is shown in Fig. 5 and appears to describe the results on the thin-film sample (sample A) below the maximum quite well.

We can now discuss the difference in the temperature dependence of $S_y$ between sample A and sample B in Fig. 5 by considering the length scale of subthermal magnons in terms of the sample thickness. Based on the suppression of LSSE at high fields, we can state that most of the LSSE signal observed at low temperatures comes from magnons at energies below $k_B T_C$. We can then extrapolate the mean free path data of Ref [23] into the range of 10 K to 40 K, and we make the reasonable assumption that these subthermal magnons will have $\ell_M$ of a few μm or above within this temperature range. Since Sample A is only 4 μm thick, we expect the subthermal magnons in this sample to be in the ballistic regime with a constant $\ell_M = 4$ μm, and therefore Eq. (9) to hold at low temperatures. This assumption of ballistic magnon transport in sample A is expected to break down once the temperature becomes high enough that $\ell_M$ becomes limited by the increased prevalence of magnon-phonon interactions..

In contrast to this situation, the subthermal magnons in the bulk sample B are expected to propagate diffusively at all temperatures due to the much larger YIG thickness (500 μm), since the magnon mean free path observed at 2K in Ref. [23] was only 100 μm. As in sample A, however, the magnon mean free path in sample B is also a function of temperature with a negative exponent,[23] which results in a slower $T$ dependence of the LSSE in sample A, as observed in Fig. 5. We attribute this difference to the fact that there is likely no purely ballistic magnon transport in sample B, and thus inelastic magnon scattering starts to affect the subthermal magnons starting from even the lowest temperatures examined in this study. Furthermore, we can explain the shift in peak temperature between sample A and sample B by considering that magnons in sample B (the bulk crystal) are sensitive to any mechanism that scatters magnons over a length scale shorter than 4 μm, whereas those in sample A (the 4 μm film) are not. In other words, boundary scattering persists as more important than diffuse inelastic magnon-phonon or magnon-magnon[39] scattering for magnons up to $T_C$ ~ 40 K in sample A, whereas the inelastic processes are already significant above 45 K in sample B. As a result, the maximum in $S_y$ in sample B appears at 70 K, a much lower temperature than in sample A ($T_{avg}$ = 180 K).

It is noted that the $S_y$ of sample A is larger than that of sample B above 80 K and becomes smaller below 80 K because of the faster $T$ dependence. Kehlberger et al.[25] demonstrated that the LSSE signal increases as the thickness of YIG increases up to 100 nm and starts to saturate for larger thicknesses at room temperature, while Rezende et al.[18] observed no difference in LSSE between 8 μm thick and 1 mm thick YIG samples under the same $\nabla T_z$. Both data are consistent with each other in the sense that 8 μm is likely to already be larger than the mean free path of subthermal magnons at room temperature (~ 100 nm), which can be estimated from the data in Ref. [23]. The same argument does not apply to our result, however, as the magnitude of $S_y$ is quite different in sample A and sample B at room temperature. Indeed, we find that the magnitude of $S_y$ can be a quite complex problem because the magnitude of the voltage induced by the LSSE is sensitive to the Pt/YIG interface condition.[40,41,42] Furthermore, because of the different origins of sample A and sample B in our study, we cannot exclude the possibility that the Pt/YIG interface condition varies between the two samples, which may at least partially explain the observed difference in the absolute magnitude of $S_y$.

## CONCLUSIONS

In summary, we present magnetic field and temperature dependent data of the longitudinal Spin-Seebeck effect on the YIG/Pt system on a 4 μm thick film and on a bulk YIG. From the freeze-out of the LSSE signal in high-field data at different temperatures below 300 K, we conclude that subthermal magnons (defined as those with energies below a critical energy $k_B T_C$ with $T_C$ of the order of 40 K) must play a more important role in the LSSE than most magnons present at thermal equilibrium at $T > T_C$. We also attribute the temperature-dependence of the LSSE predominantly to that of the thermally-driven magnon flux, for which we develop a new model that takes into account the departure of the magnon dispersion relations from purely parabolic to a relation that more closely resembles that measured by neutron diffraction. Finally, we argue that it is this non-parabolic dispersion that is the reason behind the observed decrease in LSSE signal with decreasing temperature below the maximum, which deviates from the expected $T^1$ law for parabolic magnons when their mean free path is constant.

## ACKNOWLEDGMENTS

The work is supported as part of the ARO MURI under award number W911NF-14-1-0016, US AFOSR MURI under award number FA9550-10-1-0533 (HJ), NSF grants CBET-1133589 and DMR 1420451, and NSF MRSEC (SB). The thin-film sample was kindly supplied by Professors E. Saitoh and K. Uchida, Tohoku University, Sendai, Japan, whom we also acknowledge for useful conversations.

**Figure Captions**

FIG. 1. (Color online) Experimental setup for LSSE measurement. (a) Experimental setup on a liquid helium cryostat used for both sample A and B. (b) Schematic illustration of the measurement geometry for sample A. The same geometry applies to sample B except that the GGG substrate is absent in sample B. The sample dimensions are not drawn to scale.

FIG. 2. (Color online) Low field LSSE measurement. (a) – (c) Transverse voltage $V_y$ as a function of magnetic field $H_x$ (a,b), and $\Delta V_y$ versus applied temperature difference $\Delta T_z$ (c) for sample A. (d) – (f) $V_y$ as a function of $H_x$ (d,e), and $\Delta V_y$ versus $\Delta T_z$ (f) for sample B.

FIG. 3. (Color online) High field LSSE measurement on sample B. (a), (b) $V_y / \Delta T_z$ as a function of $H_x$ measured up to $H_x = \pm 70$ kOe ("high field") at $T_{avg} = 296$ K (a) and at $T_{avg} = 10$ K (b). (c), (d) Magnification of low field ($H_x = \pm 3$ kOe) results at $T_{avg} = 296$ K (c) and at $T_{avg} = 10$ K (d).

FIG. 4. (Color online) (a) $T$ dependence of $\Delta S_y / S_{y,max}$ for sample B. Here, $\Delta S_y / S_{y,max} \equiv (S_{y,max} - S_{y,70kOe}) / S_{y,max}$ with $S_{y,max}$ and $S_{y,70kOe}$ defined as in the text. $\Delta S_y / S_{y,max} = 1$ indicates the fully suppressed $S_y$ at $H = \pm 70$ kOe (i.e. $S_{y,70kOe} = 0$). The inset shows the same data extended to 300 K in log scale. $T_{avg}$ denotes the average sample temperature. (b) Schematic diagram of magnon dispersion of YIG below 0.6 THz (30 K) per Ref [27]. The red arrow indicates opening of a magnon gap equivalent to ~ 9 K when $H = \pm 70$ kOe is applied. The red dashed lines mark the temperature (~ 15 K) below which significant increase in $\Delta S_y / S_{y,max}$ starts to occur.

FIG. 5. (Color online) $T$ dependence of the spin Seebeck coefficient $S_y$ for the 4 μm film sample A (green diamond) and the bulk sample B (orange circle). The dashed lines represent a power law for the bulk sample, and values calculated from the model (Eq. 9) in the text for the thin-film sample, renormalized to go through the data.

**Figure 1 (one column)**

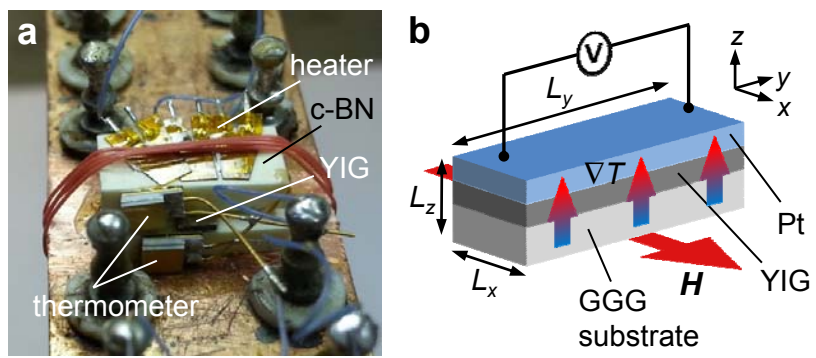

**Figure 2 (two columns)**

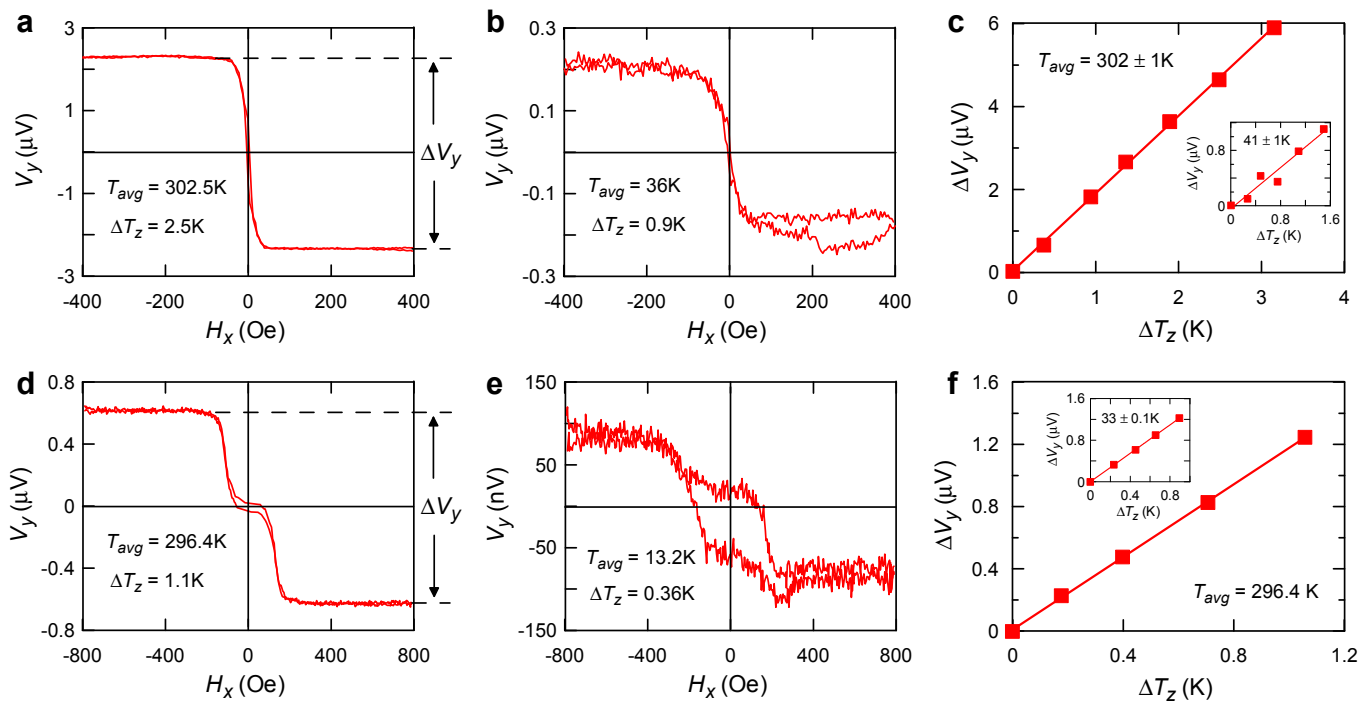

**Figure 3 (one column)**

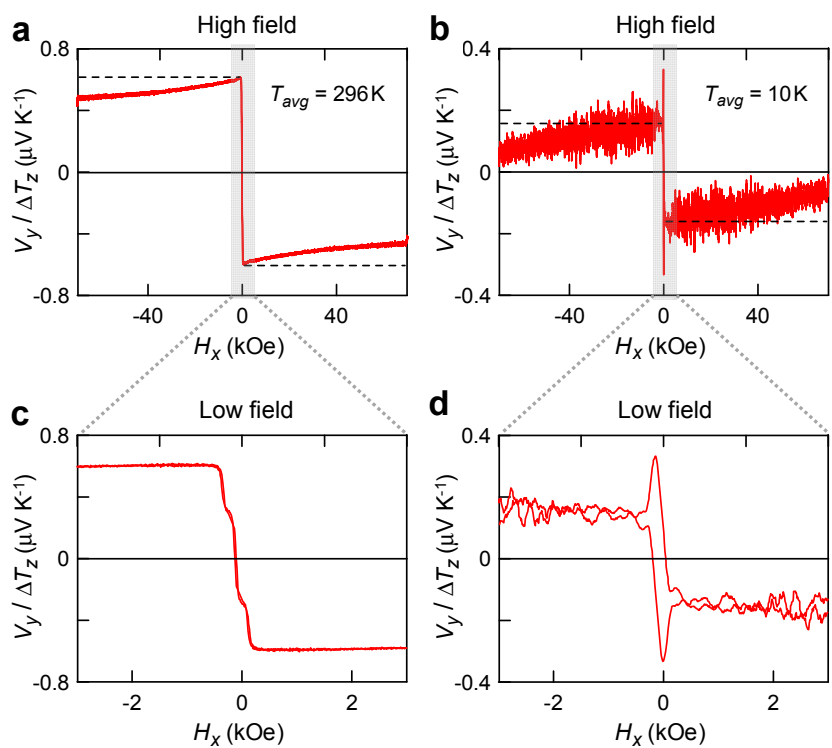

**Figure 4 (one column)**

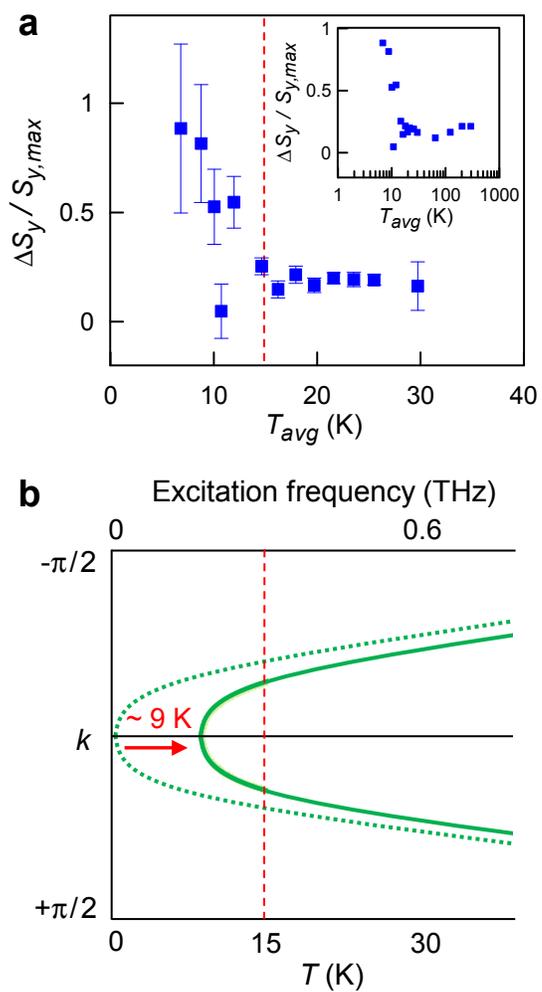

**Figure 5 (one column)**

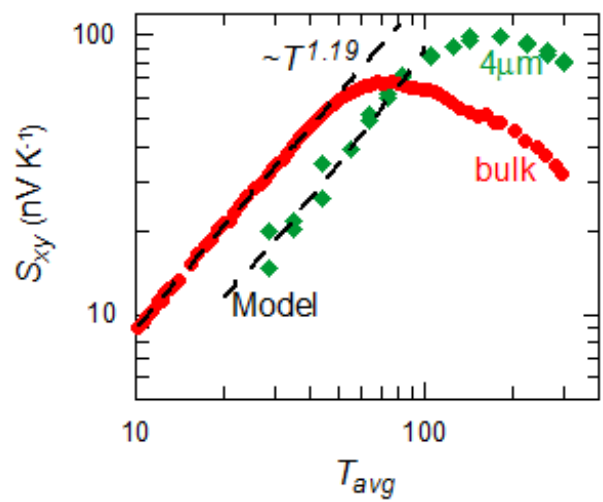
23